\documentclass[nofootinbib,floatfix]{revtex4}
\usepackage{graphicx}
\usepackage{epstopdf}

\newcommand{\tfo}{T_{\rm f.o.}}
\newcommand{\trh}{T_{\rm RH}}
\newcommand{\sigmav}{{\langle\sigma v\rangle}}

\usepackage{color}

\begin{document}

\title{The effect of a  late decaying scalar on the neutralino relic density}
\author{
\mbox{Graciela Gelmini$^{1}$\rlap,}
\mbox{Paolo Gondolo$^{2}$\rlap,}
\mbox{Adrian Soldatenko$^{1}$\rlap,}
\mbox{Carlos E. Yaguna$^{1}$}
}
\affiliation{
\mbox{$^1$  Department of Physics and Astronomy, UCLA,
 475 Portola Plaza, Los Angeles, CA 90095, USA}
\mbox{$^2$ Department of Physics, University of Utah,
   115 S 1400 E \# 201, Salt Lake City, UT 84112, USA }
\\
{\tt gelmini,asold,yaguna@physics.ucla.edu},
{\tt paolo@physics.utah.edu}}

\begin{abstract} \noindent
 If the energy density of the Universe before nucleosynthesis is  dominated by a scalar field $\phi$ that decays and reheats the plasma to a temperature $T_{RH}$ smaller than the standard neutralino freeze out temperature, the neutralino relic density differs from its standard value. In this case, the relic density depends on two additional parameters: $T_{RH}$, and the number of neutralinos produced per $\phi$ decay per unit mass of the $\phi$ field.
In this paper, we numerically study  the neutralino relic density as a function of these reheating parameters within minimal supersymmetric standard models and
show that the dark matter constraint can  almost always be satisfied.
\end{abstract}

\maketitle

\section{Motivation}

The neutralino is considered  a good dark matter candidate, one that naturally yields the required relic density. Recently, however, it has been recognized that, owing to the experimental constraints and to the increased precision in the determination of the dark matter content of the Universe, the agreement between the observed relic density,
 $\Omega_{\rm cdm} h^2 =
0.113\pm0.009 $ \cite{wmap}\footnote{There is no qualitative change in our conclusions if we would use the recent more precise value  $\Omega_{\rm cdm} h^2 =
0.109^{+0.003}_{-0.006} $, obtained for a $\Lambda$CDM model with scale-invariant primordial perturbation spectrum through a global fit of cosmic microwave background, supernovae, and large scale structure data in \cite{wmap2}.} and the relic density predicted with standard cosmological assumptions, 
 is far from being a generic feature of supersymmetric models. In fact, models with  bino-like neutralinos tend to overproduce them, and special mechanisms such as coannihilations or  resonant annihilations are required to suppress the relic density down to the observed range. In contrast, models with  higgsino- or  wino-like neutralinos usually give too small a relic density and, to compensate for it, large neutralino masses ($m_\chi>1\,\mathrm{TeV}$) are needed. Non-standard cosmologies and in particular models with low reheating temperatures provide a plausible solution to this problem. 
 These include models with
moduli decay~\cite{ Moroi:1999zb}, Q-ball decay \cite{Fujii}, and
thermal inflation~\cite{Lazarides}. In all of these models there is a
late episode of entropy production and  non-thermal
production of the LSP in particle decays is possible.

We concentrate on cosmological models in which the early Universe is dominated by the energy density of a scalar field that  after some time decays giving rise to the radiation dominated era. The decay of the scalar field into light degrees of freedom and their subsequent thermalization -the reheating process- leaves the Universe at a temperature $T_{RH}$ known as the reheating temperature. If, as assumed in the standard scenario, $T_{RH}$ is larger than the neutralino freeze out temperature ($T_{\rm f.o.}\simeq m_\chi/20$), the neutralino relic density is insensitive to its value. But, because we have no physical evidence of the radiation dominated Universe before big-bang nucleosynthesis, $T_{RH}$ should be considered as a cosmological parameter that can take any value above a few $\mathrm{MeV}$ \cite{Kawasaki:2000en,Hannestad:2004px}.

The existence of a weakly coupled scalar field that dominates the Universe during the process of neutralino production and freeze out may affect the relic density in several ways. It modifies the temperature-scale factor and the temperature-expansion rate relations \cite{McDonald,  Chung, Giudice:2000ex} that determine the freeze out condition. It dilutes the neutralino thermal abundance by increasing the entropy of the Universe  with its decay into radiation. It may also increase the neutralino relic density by decaying into supersymmetric particles which in turn decay into neutralinos. As a result, the neutralino relic density may lie above or below its standard value depending on the reheating parameters.

 Several papers dealing with the neutralino relic density in the presence of a decaying scalar field already exist in the literature \cite{Moroi:1999zb, Chung, Giudice:2000ex,  kamionkowski-turner, Moroi, Drees, Khalil, Fornengo:2002db, Pallis:2004yy, endo, kohri}. None of  these, however, presents a systematic study of the combined effect of low reheating temperatures and non-thermal production, nor their implications within general supersymmetric  models. The systematic study of thermal and non-thermal production mechanisms was introduced in Ref.~\cite{GG}. Here we study the implications within the Minimal Supersymmetric Standard Model (MSSM).
  After reviewing the standard scenario, the equations that describe the reheating process will be introduced and numerically solved for different sets of parameters. We will elucidate the differences with respect to the standard computation and will investigate the effects of the reheating parameters on the neutralino relic density.  Then, we will compute $\Omega_\chi h^2$ in generic  MSSM supersymmetric models for  different sets of the reheating parameters. Finally, a brief discussion  of our results as well as our conclusions will be presented. 

\section{The standard scenario}
In the standard scenario, it is assumed that neutralinos are produced by scattering from the thermal plasma in a radiation dominated Universe. The neutralino relic density is then determined by the Boltzmann equation for the neutralino number density ($n$) and the law of entropy   conservation:
\begin{eqnarray}
\frac{dn}{dt}&=&-3Hn-\left<\sigma v\right>(n^2-n_{eq}^2)\,,\label{eq:first}\\
\frac{ds}{dt}&=&-3Hs\,.
\label{eq:twostd}
\end{eqnarray}
Here $t$ is time, $s$ is the entropy density, $H$ is the Hubble parameter, $n_{eq}$ is the neutralino equilibrium number density, and $\left<\sigma v\right>$ is the thermally averaged total annihilation cross section \cite{Gondolo:1990dk}. Both the expansion of the Universe and the change in number density due to annihilations and inverse annihilations are  taken into account in Eq.~(\ref{eq:first}). It is convenient to combine  these two equations into a single one for $Y=n/s$  and to use  $x=m/T$, with $T$ the photon temperature, as the independent variable instead of time:
\begin{equation}
\frac{dY}{dx}=\frac{1}{3H}\frac{ds}{dx}\left<\sigma v\right>(Y^2-Y_{eq}^2)\,.
\label{eq:std}
\end{equation}
According to the Friedman equation, the Hubble parameter is determined by the mass-energy density $\rho$ as 
\begin{equation}
H^2=\frac{8\pi}{3M_P^2} \rho\,,
\end{equation}
where $M_P=1.22\times10^{19}~{\rm GeV}$ is the Planck mass.
The energy and entropy densities are related to the photon temperature by the equations
\begin{equation}
\rho=\frac{\pi^2}{30}g_{\rm eff}(T) T^4\,,\quad s=\frac{2 \pi^2}{45}h_{\rm eff}(T) T^3
\label{eq:eff}
\end{equation}
where $g_{\rm eff}$ and $h_{\rm eff}$ are the effective degrees of freedom for the energy density and entropy density respectively. If we define the degrees of freedom parameter $g_*^{1/2}$ as
\begin{equation}
g_*^{1/2}=\frac{h_{\rm eff}}{g_{\rm eff}^{1/2}}\left(1+\frac{1}{3}\frac{T}{h_{\rm eff}}\frac{dh_{\rm eff}}{dT}\right)
\end{equation}
then Eq.~(\ref{eq:std}) can be written in the following way,
\begin{equation}
\frac{dY}{dx}=-\left(\frac{45}{\pi M_P^2}\right)^{-1/2}\frac{g_*^{1/2}m}{x^2}\left<\sigma v\right> (Y^2-Y_{eq}^2)\,.
\label{eq:stdfi}
\end{equation}
This single equation is then numerically solved with the initial condition $Y=Y_{eq}$ at $x\simeq 1$ to obtain the present neutralino abundance $Y_0$. From it, the neutralino  relic density can be computed
\begin{equation}
\Omega_{\rm std}h^2=\frac{\rho_\chi^0h^2}{\rho_c^0}=\frac{m_\chi s_0 Y_0h^2}{\rho_c^0}=2.755\times 10^8\, Y_0 m_\chi/\mathrm{GeV}\,,
\label{eq:relic}
\end{equation}
where $\rho_c^0$ and $s_0$ are the present critical density and entropy density respectively. In obtaining the numerical value in Eq.~(\ref{eq:relic}) we used  $T_0=2.726\,\mathrm{K}$ for the  present background radiation temperature and $h_{\rm eff}(T_0)=3.91$ corresponding to photons and three species of relativistic neutrinos. 

The numerical solution of Eq.~(\ref{eq:stdfi}) shows that at high temperatures $Y$ closely tracks its equilibrium value $Y_{eq}$. In fact, the interaction rate of neutralinos is strong enough to keep them in thermal and chemical equilibrium with the plasma. But as the temperature decreases, $Y_{eq}$ becomes exponentially suppressed and $Y$ is no longer able to track its equilibrium value. At the freeze out temperature ($T_{\rm f.o.}$), when the neutralino annihilation rate becomes of the order of the Hubble expansion rate, neutralino production becomes negligible and the neutralino abundance per comoving volume reaches its final value.

From this discussion follows that the freeze out temperature plays a prominent role in determining the neutralino relic density. In the standard cosmological scenario, the neutralino freeze out temperature is about $m_\chi/20$. In general, however, the freeze out temperature depends not only on the mass and interactions of the neutralino but also, through the Hubble parameter, on the content of the Universe.

\section{Low reheating temperature scenarios}

If the freeze out temperature is larger than the reheating temperature ($T_{\rm f.o.}>T_{RH}$) or equivalently if neutralinos decoupled from the plasma before the end of the reheating process, the neutralino relic density will differ from  its standard value. In that case, the dynamics of the Universe is described by the following equations
\begin{eqnarray}
\frac{d\rho_\phi}{dt}&=&-3H\rho_\phi-\Gamma_\phi \rho_\phi \label{eq:reh1}\\
\frac{dn}{dt}&=&-3Hn-\left<\sigma v\right>(n^2-n_{eq}^2)+\frac{b}{m_\phi}\Gamma_\phi \rho_\phi \label{eq:reh2}\\
\frac{ds}{dt}&=&-3Hs+\frac{\Gamma_\phi \rho_\phi}{T}
\label{eq:reh3}
\end{eqnarray}
where $m_\phi$, $\Gamma_\phi$, and $\rho_\phi$ are respectively the  mass, the decay width and the energy density of the scalar field, and $b$ is the average number of neutralinos produced per $\phi$ decay. Notice that $b$ and $m_\phi$ enter into these equations only through the ratio $b/m_\phi$ and not separately.
Finally, the Hubble parameter, $H$, receives contributions from the scalar field, Standard Model particles, and supersymmetric particles,
\begin{equation}
H^2=\frac{8\pi}{3 M_P^2}(\rho_\phi+\rho_{SM}+\rho_\chi)\,.
\label{eq:reh4}
\end{equation}

These equations take into account the evolution of the scalar field, its contribution to the entropy density,  and its possible decay into supersymmetric particles. As expected, in the limit $\rho_\phi\rightarrow 0$ they reduce to the standard scenario. In writing them, we have implicitly assumed that initially the $\phi$ field inflated the Universe to a state of negligible entropy, that in each $\phi$ decay an energy $m_\phi$ is contributed to the matter and radiation plasma, and that the decay products rapidly thermalize (i.e.\ reach kinetic, although not necessarily chemical, equilibrium).

In cosmologies with a decaying scalar, therefore, the neutralino relic density depends on two additional parameters: $b/m_\phi$, and $\Gamma_\phi$. In specific models describing the physics of the $\phi$ field, these parameters may be calculable
 (see Ref.~\cite{GG} and references therein). Our approach, however, will be completely phenomenological. We will treat $b/m_{\phi}$ and $\Gamma_{\phi}$ as free parameters subject to the constraints that the dark matter bound imposes on them. For convenience instead of $b/m_\phi$ we will sometimes use the dimensionless quantity
\begin{equation}
\eta=b\,  \left(\frac{100~\mathrm{TeV}}{m_\phi}\right)
\end{equation}
as the free parameter. Clearly, $b=\eta$ for $m_\phi=100~\mathrm{TeV}$. $\Gamma_\phi$, on the other hand,  is commonly  expressed  in terms of the parameter $T_{RH}$ -- conventionally called the reheating temperature -- defined by assuming an instantaneous conversion of the  scalar field energy density into radiation,
\begin{equation}
\Gamma_\phi=\sqrt{\frac{4 \pi^3 g_{\rm eff}(T_{RH})}{45}}\frac{T_{RH}^2}{M_P}\,.
\end{equation}

The Eqs.~\ref{eq:reh1}--\ref{eq:reh4} above assume thermal equilibrium.  As explained in Ref.~\cite{GG}, this assumption is well justified. Let us see why.
 First note that during the epoch in which the Universe
is dominated by the decaying $\phi$ field, $H$ is proportional to
$T^4$~ \cite{McDonald}. We can derive this dependence starting from the 
evolution equation of  the entropy per comoving volume $S= s a^3$, which is
eq \ref{eq:reh3} multiplied by $a^3$, namely $dS/dt= \Gamma_\phi \rho_\phi a^3/T$.
 During the oscillating $\phi$ dominated
period, $\rho_\phi a^3$ is constant, thus $a^3 \propto \rho_\phi^{-1}$. Using $H\propto \sqrt{\rho_\phi} \propto t^{-1}$ we obtain  $a^3 \propto t^2$
and writing $T \propto t^{\alpha}$ we get
 $S \simeq T^3 a^3\propto t^{3 \alpha +2}$. Substituting these expressions for $S$
 and $T$ as functions of time
 in the evolution equation for $S$,
and matching the powers of $t$ on both sides, determines  $\alpha=-(1/4)$.
Hence, during the oscillating
$\phi$ dominated epoch $H \propto t^{-1} \propto T^4$,
 and $\rho_{\phi} \propto H^2  \propto T^8$.
Since  $H \simeq \trh^2/M_P$ at $T=\trh$, it follows that $H \simeq T^4/(\trh^2
M_P)$. 

 Neutralinos are in thermal (also called kinetic) equilibrium with the radiation when their scattering rate off
relativistic particles is faster than the Hubble expansion rate,
$\Gamma^{\rm scatt} \gtrsim H$. Because  $\Gamma^{\rm scatt} \simeq n_\gamma \sigma_{\rm scatt}
\simeq T^3 \sigma_{\rm scatt}$, we see that kinetic equilibrium is
maintained at low temperatures, $T \lesssim \trh^2 M_P \sigma_{\rm
  scatt}$. This is contrary to the usual radiation-dominated scenario
in which the temperature has to be large enough for kinetic
equilibrium to be maintained. The origin of this difference is the
strong $T^4$ dependence of $H$ on the radiation temperature. Using the known
relation $\Omega_{\rm std} h^2 \simeq 10^{-10} {\rm
  ~GeV^{-2}}/\sigmav$, and taking $\sigma_{\rm scatt}$
of the same order of magnitude as the annihilation cross section $\langle\sigma
v\rangle$, gives $T \lesssim (10^6 {\rm ~MeV~}/\Omega_{\rm std} h^2)
(\trh/{\rm MeV})^2$ for kinetic equilibrium to be maintained. For the validity of Eqs.~\ref{eq:reh1}--\ref{eq:reh4}, we need to assume kinetic
equilibrium before neutralino production ceases: the right hand
side is larger than $\trh$ for $\Omega_{\rm std}h^2 \lesssim 10^6
(\trh/{\rm MeV})$, so we  safely assume that kinetic equilibrium is
reached.

To solve the system of equations (\ref{eq:reh1})-(\ref{eq:reh4}), besides $\eta$ and $T_{RH}$ we have to specify the initial conditions  for $\rho_\phi$, $n_\chi$, and $s$. Giving the value $H_I$ of the Hubble
parameter at the beginning of the $\phi$ dominated epoch amounts
to giving the initial energy density $\rho_{\phi,I}$ in the $\phi$
field, or equivalently the maximum temperature of the radiation
$T_{\rm MAX}$. Indeed, one has $H_I \simeq \rho_{\phi,I}^{1/2}/M_P
\simeq T_{\rm MAX}^4/(\trh^2 M_P)$. The latter relation can be derived from
$\rho_\phi \simeq T^8/\trh^4$ and the consideration that the maximum
energy in the radiation equals the initial (maximum) energy
$\rho_{\phi,I}$.
 At the fundamental level the initial energy density of the scalar field $\phi$ depends on the particular model describing it. In hybrid models of inflation, for instance, $\rho_{\phi,I}\simeq m_\phi^4$, whereas a much larger value is expected in  chaotic inflation $\rho_{\phi,I}\simeq m_\phi^2 M_{P}^2$. From a physical point of view it is not surprising that the relic density is sensitive to the initial $\phi$ energy density, for it determines the available energy of the Universe. If $\rho_{\phi,I}$ is too small, the Universe can never get hot enough to thermally produce neutralinos and the relic density is typically suppressed. 
 If the neutralino reaches chemical equilibrium, it is clear that its
final density does not depend of the initial conditions. 
An approximate condition for reaching chemical equilibrium is \cite{Giudice:2000ex}
$\sigmav \gtrsim 10^{-9} {\rm GeV}^{-2}$ $(m_\chi/100{\rm
  GeV}) (\trh/{\rm MeV})^{-2}$. Even without reaching chemical equilibrium, the
neutralino density is insensitive to the initial conditions provided the
maximum temperature of the radiation $T_{\rm MAX} \gtrsim
m_\chi$ \cite{Giudice:2000ex}.

Regarding the other two initial conditions, we assume that at an arbitrary reference temperature $T_i$ the initial entropy is given by Eq.~(\ref{eq:eff}), and that the neutralino number density is negligible, $n_i=0$. We explicitly checked that when chemical equilibrium is reached, using different values for $T_i$ and $n_i$ does not modify the neutralino relic density. Thus, our results will all be independent on the initial conditions.

At early times, when $t\Gamma_\phi\ll 1$ and the $\phi$ field is oscillating around the minimum of its potential, the scalar field energy density per comoving volume $\rho_\phi a^3$ is essentially constant and useful analytical results can be derived from eqns (\ref{eq:reh1})-(\ref{eq:reh4}). The temperature and the scale factor are related by $T\propto a^{-3/8}$ whereas the temperature and the expansion rate are linked by $H\propto T^4$, marking the departure from the standard cosmological scenario --where $T\propto a^{-1}$ and $H\propto T^2$.  This is not the only relevant departure, though. From a qualitative point of view, at least three  important differences between the standard cosmology and the scenario with a decaying scalar field can be identified. They are
\begin{itemize}
\item During freeze out the Universe may  not be dominated by radiation. Initially, the energy density is dominated by the scalar field $\phi$ and only at the end of the reheating process the Universe becomes radiation dominated. 
\item The entropy density of the Universe is not conserved but increases during the reheating era due to the scalar field decay.
\item Neutralinos could be produced non-thermally in $\phi$ decays. This production mechanism could even be the dominant source of neutralinos. 
\end{itemize}

 In Ref.~\cite{GG} the solutions to Eqs.~(\ref{eq:reh1}--\ref{eq:reh4}) were obtained analytically (as well as numerically).  There are 
four different  cases which result from the different ways in which the density
$\Omega_\chi h^2$ depends on $\trh$:

Case (1), thermal production without
chemical equilibrium. In this case  $\Omega_{\chi} \propto \trh^7$.
The relic density was estimated
in Ref.~\cite{Chung}:
\begin{equation} 
\frac{ \Omega_\chi}{\Omega_{\rm cdm}} \simeq 
\frac{\sigmav}{10^{-16}~{\rm GeV}^{-2}}
\left(\frac{100{\rm GeV}}{m_\chi}\right)^{5}
\left(\frac{\trh}{{\rm GeV}}\right)^7
\left(\frac{10}{g_\star}\right)^{3/2}.
\end{equation}

Case (2), thermal production with chemical equilibrium. In this case
 $\Omega_{\chi} \propto
\trh^4$.
The neutralino freezes out while the universe is
dominated by the $\phi$ field. Its freeze-out density is larger than
usual, but it is diluted by entropy production from
$\phi$ decays. The new freeze-out temperature $\tfo^{\rm NEW}$ is determined
by solving $n \sigmav \simeq H$ at $T=$ $\tfo^{\rm NEW}$. Using the relations
between $H$, $a$, and $T$ in the decaying-$\phi$ dominated Universe,
one finds \cite{McDonald,Giudice:2000ex} 
\begin{equation}
\Omega_\chi \simeq
\trh^3\tfo(\tfo^{\rm NEW})^{-4} \Omega_{\rm std}. 
\label{eq:case2}
\end{equation}
Our numerical results indicate a slope closer to $\trh^4$, 
at least partially  due to the change in $\tfo^{\rm NEW}$ (see Fig.~ \ref{fig:evol} below)

 Case (3), non-thermal production
without chemical equilibrium. Here $\Omega_{\chi} \propto \trh$.
Non-thermal production is not compensated by annihilation. 
The production of neutralinos is purely
non-thermal and the relic density depends on $\eta$. It can be estimated
analytically as follows. For each superpartner produced, at least one LSP will
remain at the end of a chain of decays (due to $R$-parity
conservation), and thus $n_{\chi} \simeq bn_\phi$. Here $n_\phi =
\rho_\phi/m_\phi$.  At the time of $\phi$-decay
$ \rho_{\chi} \simeq m_\chi b \rho_\phi/m_\phi \simeq \trh^4$, and the
entropy  is $s \simeq \trh^3$. Hence $
\rho_\phi/s\simeq \trh$ and $Y_0=Y_{\rm decay}\simeq b \trh/m_\phi$. It follows that,
\begin{equation} 
\frac{\Omega_{\chi}}{\Omega_{\rm cdm}} \simeq 2 \times 10^3 \eta \left(\frac{m_\chi}{\rm
100~GeV}\right) \left(\frac{\trh}{\rm MeV}\right)
\label{nonthermal}
\end{equation}

Case (4), non-thermal production with
chemical equilibrium. In this case $\Omega \propto \trh^{-1}$.
Annihilation compensates for the
non-thermal production of neutralinos until the non-thermal production ceases at
$T=\trh$. The condition for determining the relic density is
$\Gamma^{\rm ann} \simeq \Gamma_{\phi}$ at $T=\trh$. This leads to $ Y_0
\simeq Y_{\rm RH} \simeq \Gamma_\phi/(s_{\rm RH} \langle\sigma v
\rangle) \simeq 1/(\trh M_P \sigmav)$. From here it follows that 
\begin{equation}
\Omega_\chi \simeq (\tfo/\trh) \Omega_{\rm std}. 
\label{eq:case4}
\end{equation}

\begin{figure}[ht]
\includegraphics[scale=0.5]{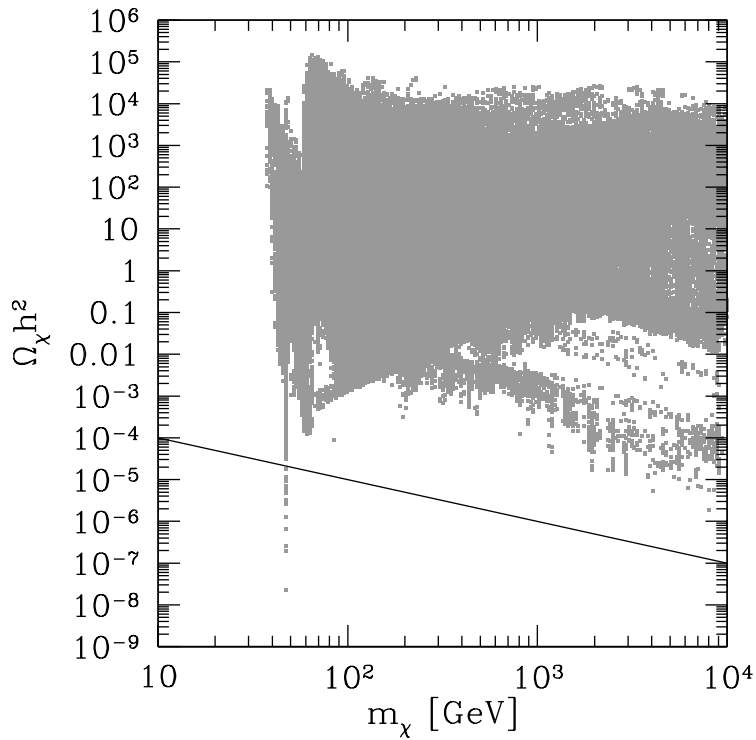}
\caption{Range of values of the standard relic density for mSUGRA models. Only those with $\Omega_{\rm std} \lesssim 10^{-5} (100 {\rm GeV}/m_\chi)$,  i.e. those below the slanted line, cannot be brought to have the total density of dark matter in low reheating temperature scenarios.}
\label{fig:Omega}
\end{figure}

In Ref.\cite{GG}, it was concluded that only neutralinos whose standard relic density is $\Omega_{\rm std} \lesssim 10^{-5} (100 {\rm GeV}/m_\chi)$ cannot be
brought to $\Omega_{\rm cdm}$, independently of $\eta$, while all neutralinos with larger $\Omega_{\rm std}$ can be made to account for all of cold dark matter within these scenarios. In order to have an idea of how many neutralino models cannot be brought to have the full dark matter density, Fig.~\ref{fig:Omega} shows the range of values of  $\Omega_{\rm std}$ for a few million mSUGRA models (data courtesy of Ted Baltz). One can see that very few of them, namely only some with $m_\chi \simeq m_Z/2$ for which neutralinos annihilate resonantly through a Z boson, have $\Omega_{\rm std} \lesssim 10^{-5} (100 {\rm GeV}/m_\chi)$, i.e. are below the slanted line in the figures.

\begin{figure}[tb]
\includegraphics[scale=0.5]{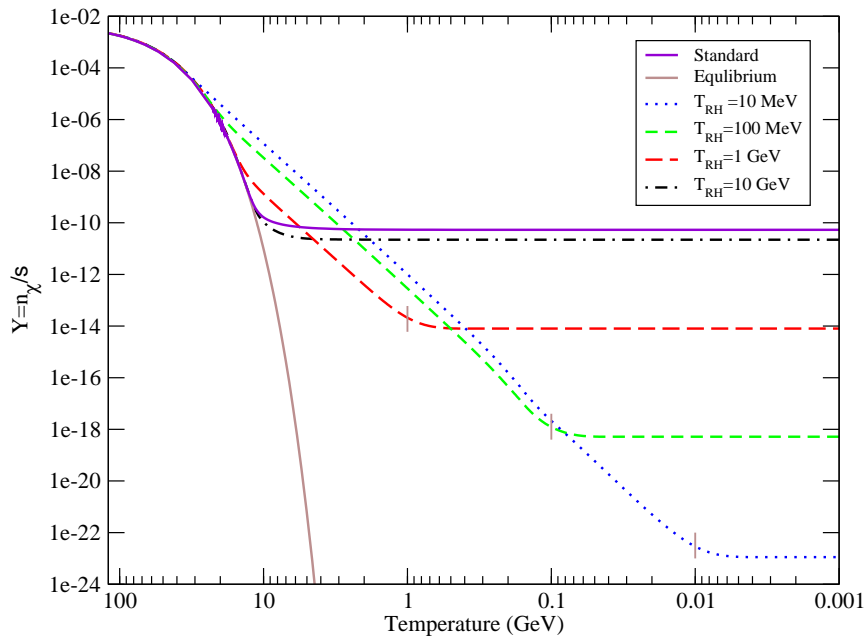}
\caption{The evolution of the neutralino abundance for different values of $T_{RH}$ and $\eta=0$ (i.e. with only thermal production). In this figure we use the mSUGRA parameters $M_{1/2}=m_0=600\,\mathrm{GeV}$, $A_0=0$, $\tan\beta=10$, and $\mu>0$. The neutralino mass is $m_\chi=246 \,\mathrm{GeV}$.  The standard relic density is $ \Omega_{\rm std} h^2\simeq 3.6$.}
\label{fig:evol}
\end{figure}

\begin{figure}[tb]
\includegraphics[scale=0.45]{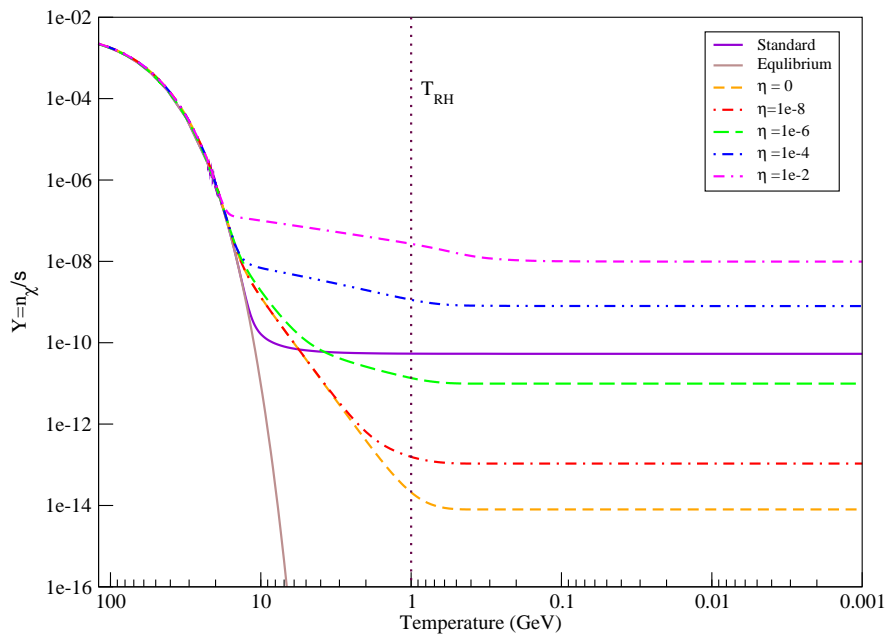}
\caption{The evolution of the neutralino abundance for $T_{RH}=1\,\mathrm{GeV}$ and several values of $\eta$. Minimal SUGRA parameters as in
Fig.~\ref{fig:evol}.}
\label{fig:evol2}
\end{figure}

In the following  we  investigate quantitatively how  the reheating process modifies the evolution of the neutralino abundance during the reheating era. To numerically integrate the equations describing the reheating process, we  modified the DarkSUSY \cite{Gondolo:2004sc} code so that the three equations (\ref{eq:reh1})-(\ref{eq:reh3}) are solved simultaneously. The advantage of using a program like DarkSUSY is twofold. On the one hand, it provides an efficient and precise algorithm to solve differential equations like (\ref{eq:reh2}). On the other hand, it automatically computes the total annihilation cross section $\left<\sigma v\right>$ in terms of supersymmetric parameters.

To begin our analysis, let us consider thermal production only ($\eta=0$) and let us see how low reheating temperatures affect the neutralino relic density. In Fig.~\ref{fig:evol} we plot the evolution of the neutralino density $Y=n/s$ as a function of $T$ for a particular choice of mSUGRA parameters: $M_{1/2}=m_0=600~\mathrm{GeV}$, $A_0=0$, $\tan\beta=10$, and $\mu>0$. For this choice, $\Omega_{\rm std} h^2 = 3.6$, thus we are examining the case of a slightly overdense neutralino. The solid lines show the equilibrium density and the neutralino density in the standard scenario. The other lines show the effect of different reheating temperatures between $10~\mathrm{GeV}$ and $10~\mathrm{MeV}$. Note that the value of the parameter $T_{RH}$, shown by  a short vertical line in each curve, coincides well with the actual reheating temperature (at which the curve changes slope, indicating that the oscillating $\phi$ dominated era finishes and the radiation dominated era begins).  We are in our case (2), thermal production with chemical equilibrium, for which $\Omega_{\chi} h^2 \propto T_{RH}^4$, as verified in the figure. The curve for $T_{RH}=10 ~\mathrm{GeV}$ closely tracks the standard one,  yielding an abundance just below the standard prediction. As we move to smaller reheating temperatures, the freeze out occurs earlier,  at a temperature $\tfo^{\rm NEW} > \tfo $ and so the neutralino abundance at freeze out  is larger than in the standard scenario. But, between $\tfo^{\rm NEW} $ and $T_{RH}$ the neutralino abundance is diluted away by the entropy released during the decay of the scalar field. In this regime, represented by the descending straight lines, $Y\propto T^5$. Indeed, $n_\chi a^3$ (with $a$ the scale factor) remains constant whereas from $a\propto T^{-8/3}$ follows that $s a^3$ decreases as $T^{-5}$; hence $ Y = n_\chi/s \propto T^5$. Because the entropy dilution effect is dominant,  the neutralino relic abundance ends below the standard prediction and the smaller the reheating temperature, the smaller the relic density. Thus, if neutralinos are produced thermally, a low reheating temperature entails a suppression of the relic density with respect to its standard value. 

\begin{figure}[tb]
\includegraphics[scale=0.45]{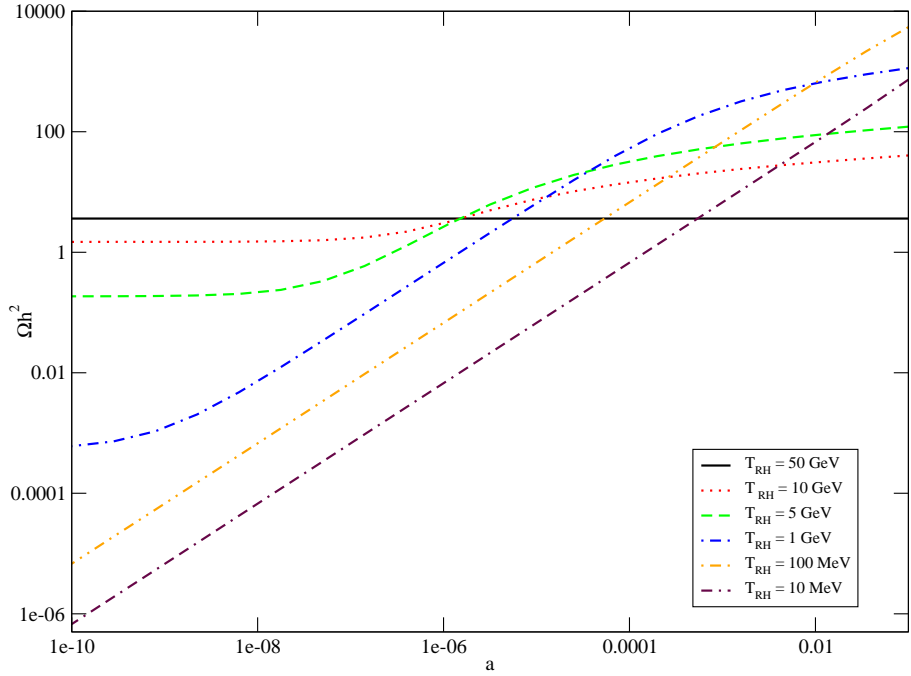}
\caption{The neutralino relic density as a function of $\eta$ for several values of  $T_{RH}$.  Minimal SUGRA parameters as in
fig \ref{fig:evol}.}
\label{fig:grb}
\end{figure}

\begin{figure}[tb]
\includegraphics[scale=0.45]{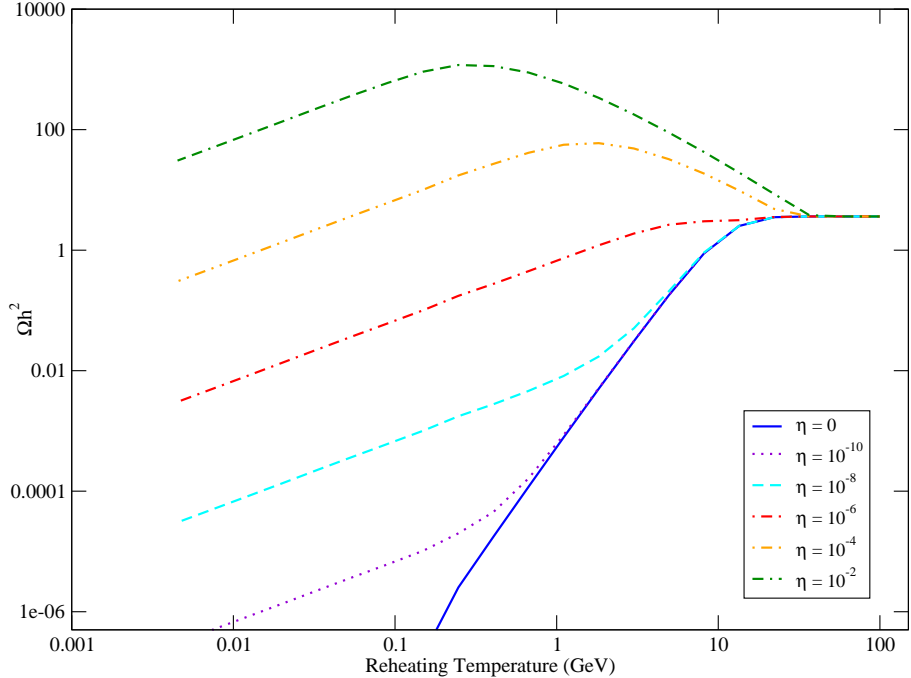}
\caption{The neutralino relic density as a function of $T_{RH}$ for different values of $\eta$.  Minimal SUGRA parameters as in
fig \ref{fig:evol}.}
\label{fig:grt}
\end{figure}

\begin{figure}[tb]
\includegraphics[scale=0.5]{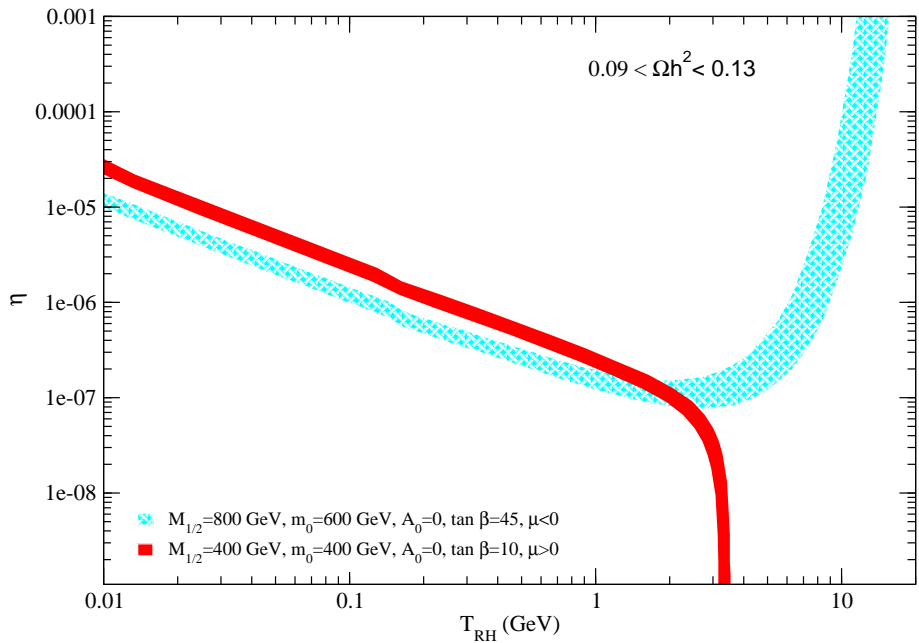}
\caption{Regions in the plane $(T_{RH},\eta)$ compatible with the WMAP range for two different supersymmetric spectra. The solid region corresponds to $m_0=M_{1/2}=400 ~\mathrm{GeV}$, $\tan\beta=10$, $A_0=0$, and $\mu>0$, parameters which yield
 $\Omega_\chi h^2=1.8$ and $m_\chi=160$ GeV . The hatched region corresponds to $m_0=600 ~\mathrm{GeV}$, $M_{1/2}=800 ~\mathrm{GeV}$, $A_0=0$, $\tan\beta=45$, and $\mu<0$, parameters which yield $\Omega_\chi h^2=0.02$ and $m_\chi=333$ GeV.}
\label{fig:btr}
\end{figure}

Non-thermal production of neutralinos, however, can compensate for such suppression. In Fig.~\ref{fig:evol2}, we show the evolution of the neutralino abundance  for  $T_{RH}=1 ~\mathrm{GeV}$ and different values of $\eta$. The dependence of $Y$ on the temperature between $T_{\rm f.o.}$ and $T_{RH}$ is now different due to the production of neutralinos in the $\phi$ decay. As we go from $0$ to larger values of $\eta$, the neutralino abundance increases and at $\eta\simeq 10^{-5}$ it becomes larger than the standard prediction.  We are in our case (3), non-thermal production
without chemical equilibrium, in which production is not compensated by annihilation. 
In Eq.~\ref{nonthermal} we see that for fixed $T_{RH}$
the final relic abundance increases with $\eta$,  $\Omega_\chi \propto \eta$. Setting $\Omega_\chi = \Omega_{\rm cdm}$, we find the critical value of $\eta$ above which the relic abundance is 
larger than its standard value. Replacing $T_{RH}=1~\mathrm{GeV}$,  $m_\chi =246~\mathrm{GeV}$ and $\Omega_{\rm cdm} h^2=0.11$, we find that $\Omega_\chi h^2$ becomes larger than $ \Omega_{\rm std} h^2\simeq 3.6$ for  the critical value $\eta_c\simeq 0.7 \times10^{-5}$.
 For larger values of $\eta$, the density of neutralinos is large enough for annihilations to become important.  We are then in our case (4), non-thermal production with chemical equilibrium, and Eq.~\ref{eq:case4} gives the final abundance. Equating Eqs.~\ref{nonthermal} and \ref{eq:case4} gives the value of $\eta$ at the cross-over, which is 
\begin{equation}
\eta_{c-o} \simeq  2.5 \left( \frac{\Omega_{\rm std}}{\Omega_{\rm cdm}} \right) \left( \frac{{\rm MeV}}{T_{RH}} \right)^2.
\label{eq:crossover}
\end{equation}
 This cross-over value of $\eta$ is about 0.8 $\times 10^{-4}$ for the example we are considering. Thus the curve with $\eta =10^{-2}$ in Fig.~\ref{fig:evol2} corresponds to
case (4).
For large values of $\eta$, the non-thermal production of neutralinos is efficient and $Y\propto \eta T$  for $T>T_{RH}$  (as follows from  $N_\chi = n_\chi a^3\propto \eta t\propto \eta T^{-4}$). For $\eta=10^{-2}$, for instance, the relic density is more than two orders of magnitude larger than the standard one. Hence, if neutralinos are produced non-thermally a low reheating temperature can yield a relic density below or above the standard result, depending on the values of $T_{RH}$ and $\eta$.

What can be constrained with the observations today, however, is not the evolution of the neutralino abundance but just its asymptotic value --the neutralino relic density. In
 Fig.~\ref{fig:grb} we show $\Omega_\chi h^2$ as a function of $\eta$ for different values of $T_{RH}$. In the figure, $T_{\rm f.o.} \simeq 12~\mathrm{GeV}$. For $T_{RH}=50~\mathrm{GeV}$ (solid line), the relic density is independent of $b$ and equal to its standard value indicating that the freeze out took place after the reheating era, in a radiation dominated Universe. For smaller values of $T_{RH}$, however,
 $\Omega_\chi h^2$ does depend on $\eta$. This is the regime of case (3), Eq.~\ref{nonthermal},
  in which non-thermal production is not compensated by annihilation. As mentioned above, setting $\Omega_\chi = \Omega_{\rm cdm}$ in Eq.~\ref{nonthermal} one finds for each value of $T_{RH}$ the critical value of $\eta$ above which the relic abundance is 
larger than its standard value. For the particular model in the figure this value is $\eta_c\simeq 0.7 \times 10^{-2} ({\rm MeV}/ T_{RH})$.
Eq.~\ref{eq:crossover} gives the cross-over value $\eta_{c-o}\simeq 0.8 \times 10^{2} ({\rm MeV}/ T_{RH})^2$ at which annihilation becomes important, and for larger values of $\eta$ the relic density is given by Eq.~\ref{eq:case4} (case (4)). Notice that  $\eta_c\simeq \eta_{c-o}$ for  $T_{RH}\simeq 10  \mathrm{GeV}$.

It is clear from  Fig.~\ref{fig:grb} that the relic density is below the standard result for small $\eta$, it increases with $\eta$ and at a certain point, $\eta=\eta_c$, coincides with  the standard one; for $\eta>\eta_c$, the relic density is larger than in the usual cosmology. The value of $\eta_c$ is inversely proportional to $T_{RH}$ at low reheating temperatures, $T_{RH} \leq 1 \mathrm{GeV}$ in the figure.  In this regime,  the relic density becomes a straight line signaling dominant non-thermal production of neutralinos without chemical equilibrium (case 3, Eq.~\ref{nonthermal}, $\Omega_\chi h^2\propto \eta$).

In Fig.~\ref{fig:grt}, the relic density is shown as a function of $T_{RH}$ for different values of $\eta$.  This figure is similar
to Fig.~1 of Ref.~\cite{GG}. We observe that  at small $T_{RH}$, $T_{RH} < 1 \mathrm{GeV}$ in the figure, the relic density  is proportional to $T_{RH}$.  This is the regime of 
non-thermal production without chemical equilibrium, in which $\Omega_\chi h^2\propto \eta T_{RH}$ (case (3), Eq.~\ref{nonthermal}).
As  $T_{RH}$ increases above 1GeV there are in the figure some descending and some ascending curves. The descending curves correspond to non-thermal production with chemical equilibrium, in which $\Omega_\chi h^2 \propto 1/T_{RH}$
 (case (4), Eq.~\ref{eq:case4}). The ascending curves correspond to thermal production with chemical equilibrium, where
  $\Omega_\chi h^2 \propto T_{RH}^4$ (case (2), Eq.~\ref{eq:case2}).
  At  $T_{RH}> \tfo \simeq$ 12 GeV in the figure, $\Omega_\chi h^2$
is  independent of $\eta$ and $T_{RH}$, and equal to its standard value. As a general conclusion, depending on the particular choices  for $T_{RH}$ and $\eta$, the relic density may be larger or smaller than its standard value.

Though specific values for the supersymmetric parameters were chosen in Figs.~\ref{fig:grb},\ref{fig:grt}, they actually represent a generic situation whenever $\Omega_{\rm std}$ is small enough for chemical equilibrium to be reached at high $T_{RH}$. Taking different points in the parameter space is equivalent to rigidly translating the figures without altering their shapes.  

Another way of visualizing the effects of $T_{RH}$ and $\eta$ on the relic density is by studying the regions compatible with the WMAP range, $\Omega_{\rm cdm} h^2= 0.09 -0.13$. In Fig.~\ref{fig:btr} we show such regions in the plane $(T_{RH},\eta)$ for two different sets of mSUGRA parameters. Neither model is viable in the  standard cosmological scenario  because one overproduces dark matter (solid region, corresponding to $\Omega_{\rm std} h^2=1.8$, $m_\chi=160$ GeV)
and the other  underproduces it (hatched region, corresponding to $\Omega_{\rm std} h^2=0.02$, $m_\chi = 333$ GeV). But, as this figure illustrates, in cosmologies with low reheating temperatures the neutralinos in both models can easily account for the dark matter content of the Universe. Notice that the first model requires a reheating temperature below $4 ~\mathrm{GeV}$.  This is the value of $T_{\rm RH}$ above which thermal production dominates and the relic density becomes independent of $\eta$ and
proportional to $T_{\rm RH}^4$ (corresponding to our case (2)) until $T_{\rm RH}$ reaches the standard freeze-out temperature
$T_{\rm f.o.}\simeq m_\chi/20 \simeq 8$ GeV, at which point  the density becomes the standard relic density. Thus here $(0.11/ \Omega_{\rm std} h^2)^{1/4} T_{\rm f.o.} \simeq 
T_{\rm f.o.}/2 \simeq 4$ GeV.
The second model demands a value of $\eta$ above $10^{-7}$. It is also evident from the figure that in underabundant models for each value of $\eta$ two different reheating temperatures can be found that give the correct relic density.  In fact, in Ref.~\cite{GG} it was shown that for neutralinos with
standard densities $\Omega_{\rm cdm} \gtrsim\Omega_{\rm std} \gtrsim 10^{-5} (100 {\rm GeV}/m_\chi)$, there is
no solution for $\eta \lesssim 10^{-7} (100 {\rm GeV}/m_\chi)^2  (\Omega_{\rm cdm}/\Omega_{\rm std})$, there are two solutions for $ 10^{-7}(100 {\rm GeV}/m_\chi)^2 (\Omega_{\rm cdm}/\Omega_{\rm std}) \lesssim \eta \lesssim 10^{-4}(100 {\rm GeV}/m_\chi)$, and there is a single solution for larger values of $\eta$.

\begin{figure}[tb]
\includegraphics[scale=0.5]{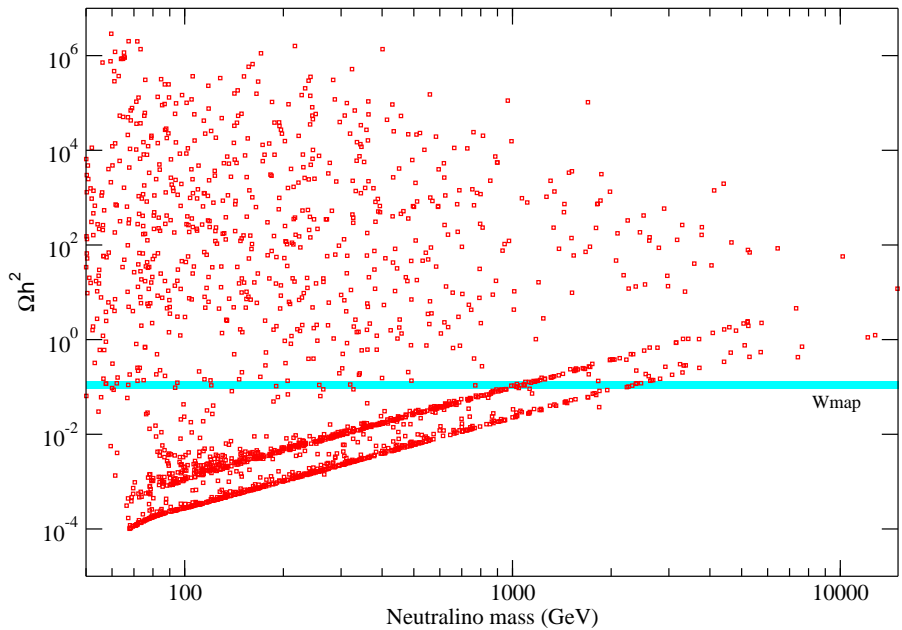}
\caption{Scatter plot of the neutralino relic density as a function of the neutralino mass in the standard scenario.}
\label{fig:std}
\end{figure}

\section{The neutralino relic density in supersymmetric models}

In the MSSM, neutralinos are linear combinations of the fermionic partners of the neutral electroweak bosons, called bino ($\tilde B^0$) and wino ($\tilde W_3^0$), and of the fermionic partners of the neutral Higgs bosons, called higgsinos ($\tilde H_u^0$, $\tilde H_d^0$). We assume that the lightest neutralino, $\chi$, is the dark matter candidate. Its composition can be parametrized as
\begin{equation}
\chi=N_{11}\tilde B^0+N_{12}\tilde W_3^0+N_{13}\tilde H_d^0+N_{14}\tilde H_u^0
\label{eq:comp}  
\end{equation}
Because the neutralino interactions are determined by its gauge content, it is useful to distinguish between bino-like, wino-like, and higgsino-like  neutralinos according to the dominant term in (\ref{eq:comp}).

\begin{figure}[tb]
\includegraphics[scale=0.45]{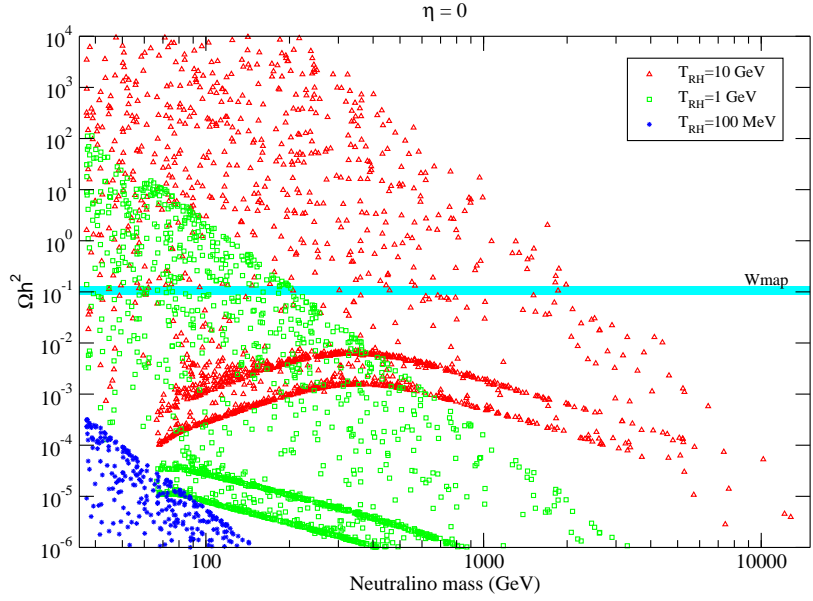}\\
\caption{Scatter plots of the relic density as a function of the neutralino mass for $\eta=0$ and different values of $T_{RH}$.
\vspace{0.5cm}}
\label{fig:b=0}
\end{figure}

 \begin{figure}[tb]
\includegraphics[scale=0.45]{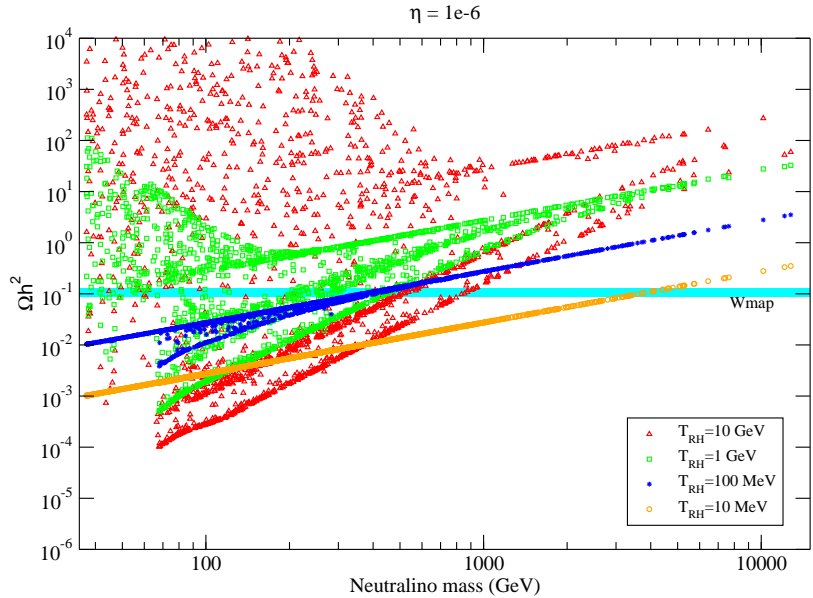}\\
\caption{As fig 8 but $\eta=10^{-6}$.}
\label{fig:b=-6}
\end{figure}

Bino-like neutralinos annihilate mainly into fermion-antifermion pairs through sfermion exchange, thus they typically have a small annihilation cross-section that translates into a  standard relic density larger than observed. Agreement with  the observed
dark matter abundance can still be achieved  in standard cosmological scenarios but only in restricted regions of the parameter space where special mechanisms such as coannihilations or resonant annihilations help reduce the relic density. Bino-like neutralinos are a generic prediction of minimal supergravity models.

Wino-like and higgsino-like neutralinos, on the other hand, annihilate mostly into gauge bosons ($W^+W^-$, $ZZ$, if kinematically allowed) through neutralino or chargino exchange; otherwise they annihilate into fermions. Through coannihilations with neutralinos and charginos of similar mass, their  standard relic density is rather small. Neutralino masses as large as $1 ~\mathrm{TeV}$ for higgsinos or  $2~\mathrm{TeV}$ for winos are required to bring their thermal density within the observed range as can be seen in Fig.~\ref{fig:std}  (see below). Wino-like and higgsino-like neutralinos can be obtained in models with non-universal gaugino masses; AMSB models, for instance, feature a wino-like neutralino.

\begin{figure}[tb]
\includegraphics[scale=0.45]{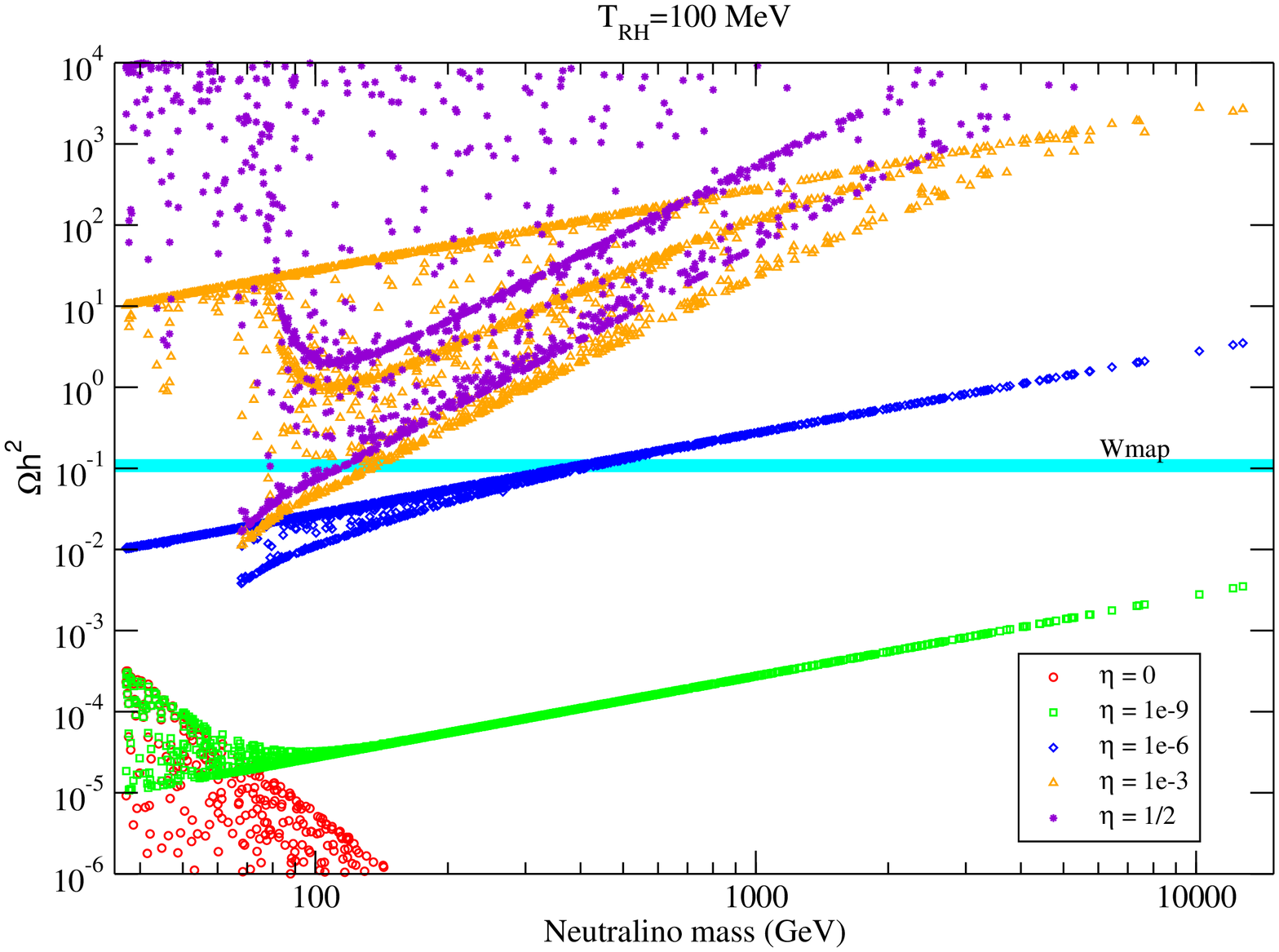}
\caption{Scatter plots of the relic density as a function of the neutralino mass for $T_{RH}=1~~\mathrm{GeV}$ and different values of $\eta$.
\vspace{0.5cm}}
\label{fig:t=-1}
\end{figure}

\begin{figure}[tb]
\includegraphics[scale=0.45]{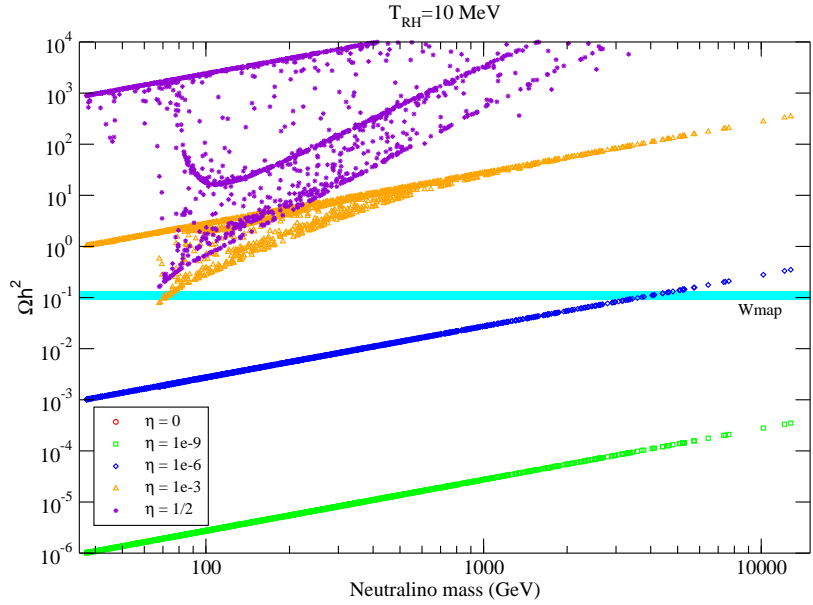}
\caption{As Fig.~10 but for $T_{RH}=10~~\mathrm{MeV}$.}
\label{fig:t=-2}
\end{figure}

Since it is our aim to study the effect of a decaying scalar on the neutralino relic density within supersymmetric models, we must consider models generic enough to allow for bino-like, wino-like, and higgsino-like neutralinos. To that end, we will examine MSSM models defined in terms of the parameter set $M_{3}$, $M_2$, $M_1$, $m_A$, $\mu$, $\tan\beta$, $m_0$, $A_t$, and $A_b$. Here $M_{i}$ are the three gaugino masses, $m_A$ is the mass of the pseudoscalar higgs boson, and $\tan\beta$ denotes the ratio $v_2/v_1$. The soft breaking scalar masses are defined through the simplifying ansatz $M_Q=M_U=M_D=M_E=M_L=m_0$ whereas the trilinear couplings are given by $A_U=\mathrm{diag(0,0,A_t)}$, $A_D=\mathrm{diag(0,0,A_b)}$, and $A_E=0$. All these parameters are defined at the weak scale. Specific realizations of supersymmetry breaking such as mSUGRA or AMSB  are particular examples of these models.

We perform a random scan of such parameter space within the following ranges
\begin{eqnarray}
10 ~\mathrm{GeV}<& M_i, m_A, \mu &< 50 ~\mathrm{TeV}\\
10 ~\mathrm{GeV}<& m_0 &<200 ~\mathrm{TeV}\\
-3 m_0<& A_t, A_b&< 3 m_0\\
1<& \tan\beta&<60
\end{eqnarray} 
A logarithmic distribution was used for $M_i$, $m_A$, $\mu$ and $m_0$, and a linear one for $A_t$, $A_b$, and $\tan\beta$; the sign of $\mu$ was randomly chosen. Accelerator constraints (as contained in DarkSUSY version 4.1~\cite{Gondolo:2004sc})  were imposed on these models. For each viable model generated in this way, we compute the neutralino relic density for different values of the reheating parameters $\eta$ and $T_{RH}$ (we computed 1,700 models for each pair of values of these two parameters and each model is a point in the scatter-plots).  In this section we present and analyze such results.

We start by showing in Fig.~\ref{fig:std} the relic density as a function of the neutralino mass in the standard cosmological scenario. The horizontal band indicates the WMAP range. From this figure we can easily tell the different kinds of neutralinos: the relic density of wino-like and higgsino-like neutralinos is determined essentially by the neutralino mass because their annihilation cross section is insensitive to the sfermion and higgs sectors. Consequently, models featuring these neutralinos are distributed along straight lines and are clearly identifiable in the figure. The line corresponding to wino-like neutralinos crosses the WMAP interval at $m_\chi\simeq 2 ~\mathrm{TeV}$, whereas that corresponding to higgsino-like neutralinos does it at $m_\chi\simeq 1 ~\mathrm{TeV}$, because winos annihilate more efficiently than higgsinos. Since the bino annihilation cross section depends on the sfermion spectrum, the relic density in models with bino-like neutralinos can vary over a large range even for a given neutralino mass. As seen in the figure, bino-like neutralinos typically yield too large a relic density in the standard cosmological  scenario.

\begin{figure}[tb]
\includegraphics[width=17cm,height=20cm]{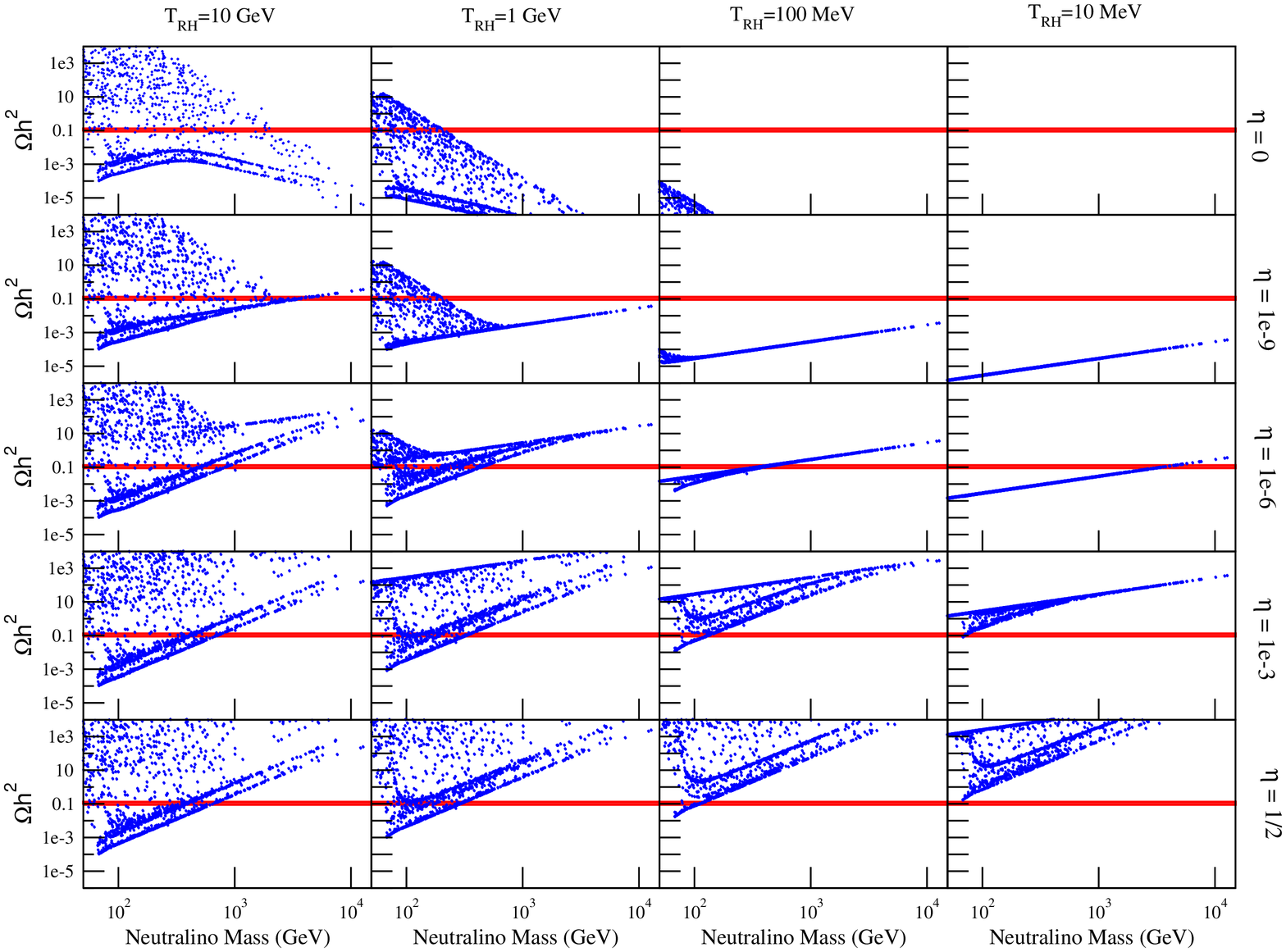}
\caption{Scatter plots of the relic density for different values of $T_{RH}$ and $\eta$.}
\label{fig:array}
\end{figure}

Let us now  consider models with a decaying scalar. A scatter plot of the relic density as a function of the neutralino mass for $\eta=0$  and different values of $T_{RH}$ is shown in Fig.~\ref{fig:b=0}. The suppression of the relic density due to small reheating temperatures is evident  in this figure.  For $T_{RH}=10 ~\mathrm{GeV}$, this effect is noticeable for large neutralino masses; the wino and higgsino lines bend downward close to $m_\chi\simeq 300 ~\mathrm{GeV}$, and heavy binos are  brought  closer to the WMAP range. For $T_{RH}=1 ~\mathrm{GeV}$ the suppression is larger and only light binos ($m_\chi<200 ~\mathrm{GeV}$) give a relic density compatible with the observations. If $T_{RH}\le 100 ~\mathrm{MeV}$ the neutralino relic density is too small to account for the dark matter of the Universe.

In general, however, low reheating temperatures  can be accompanied by non-thermal production of neutralinos, modifying  the previous results. In Fig.~\ref{fig:b=-6} we show the relic density for $\eta=10^{-6}$ and different reheating temperatures.
The straight bands with a linear dependence on $m_\chi$ shown in this and subsequent figures corresponds to our case (3) solutions, non-thermal production without chemical equilibrium, given in Eq.~\ref{nonthermal}.
 For $T_{RH}=10 ~\mathrm{MeV}$ neutralino production is entirely non-thermal  without chemical equilibrium and consequently the relic density  is proportional to the neutralino mass  (Eq.~\ref{nonthermal}). Obtaining the correct relic density requires a neutralino mass of about $4 ~\mathrm{TeV}$. If $T_{RH}=100 ~\mathrm{MeV}$ thermal effects are relevant only for light winos ($m_\chi< 200 ~\mathrm{GeV}$,  for which $T_{\rm f.o.}
 \simeq 10$ GeV $< T_{\rm RH}$); heavy winos as well as binos and higgsinos are produced non-thermally. The neutralino relic density lies within the WMAP range only for $m_\chi\simeq 400 ~\mathrm{GeV}$. For $T_{RH}=1 ~\mathrm{GeV}$, light ($m_\chi<1 ~\mathrm{TeV}$) binos, winos and higgsinos become clearly distinguishable. The correct relic abundance can be obtained at $m_\chi< 100$ GeV for binos, $m_\chi\simeq 250 ~\mathrm{GeV}$ for higgsinos, and $m_\chi\simeq 450 ~\mathrm{GeV}$ for winos. If $m_\chi> 1 ~\mathrm{TeV}$ neutralinos are produced non-thermally and  have a relic abundance well above the WMAP range. Finally, if $T_{RH}=10 ~\mathrm{GeV}$ attaining the observed dark matter density requires $m_\chi< 300 ~\mathrm{GeV}$ for binos, $m_\chi\simeq 500 ~\mathrm{GeV}$ for  higgsinos, and $m_\chi \simeq 800 ~\mathrm{GeV}$ for  winos.
 
We will now look at the effect of different values of $\eta$ for a given $T_{RH}$. In Fig.~\ref{fig:t=-1} we show the relic density as a function of the neutralino mass for $T_{RH}=100 ~\mathrm{MeV}$ and several  values of $\eta$. For $\eta=0$, the relic density is suppressed lying well below the WMAP range. For $\eta=10^{-9}$, the WMAP density can be achieved outside the mass range explored numerically along the extrapolation of the slanted straight line to higher masses; the crossing occurs at $ m_\chi \simeq 5 \times 10^5 ~\mathrm{GeV}$, as follows using Eq.~\ref{nonthermal}. With $\eta=10^{-6}$, the correct relic abundance can be achieved only for neutralino masses between $300$ and $400$ GeV. With $\eta>10^{-3}$, only a wino-like  neutralino  with $m_\chi\simeq 100-200~\mathrm{GeV}$ can account for  the dark matter.  Fig.~\ref{fig:t=-2} is as Fig.~10 but with $T_{RH}=10$ MeV. For such a small reheating temperature, the thermal production of neutralinos is negligible. The  dispersion of points discernible in this figure for $\eta=1/2$ (and for $\eta=10^{-3}$) are due to the annihilation of neutralinos produced in the decay of the $\phi$ field.  These points correspond to non-thermal production with chemical equilibrium, our case (4), eq.~\ref{eq:case4}.
It is still possible to achieve the observed relic density for $\eta$ between $10^{-6}$ and $10^{-3}$.

To summarize graphically the effects of $T_{RH}$ and $\eta$ on the relic density in generic supersymmetric models, we present in Fig.~\ref{fig:array} an array of scatter plots of the relic density for different reheating parameters. Each panel includes the same supersymmetric models  and  so the differences in the relic densities are entirely due to the effects of $\eta$ and $T_{RH}$. The reheating temperature is constant along columns while $\eta$ is constant along rows. Several conclusions can be drawn from  this figure.  The fourth row ($\eta=10^{-3}$) tells us that, if $\eta=10^{-3}$ bino-like neutralinos are disfavored due to their large relic density, and the dark matter is most easily explained with wino or higgsino-like neutralinos. From the last column ($T_{RH}=10$ MeV), we observe that if $T_{RH}=10 ~\mathrm{MeV}$, $\eta$ should be between  $10^{-6}$ and $10^{-3}$ in order to give the correct relic density. The crucial observation, however, is  that by varying $\eta$ and $T_{RH}$ we can move all the points above ($T_{RH}=10$, $\eta=1/2$) or below ($T_{RH}=100 ~\mathrm{MeV}$, $\eta=10^{-9}$) the WMAP range. For any supersymmetric model, therefore,  it is possible to find values of $\eta$ and $T_{RH}$ that give the observed value of the relic density. In other words, by choosing $\eta$ and $T_{RH}$ appropriately the dark matter bound can always be satisfied.

\begin{figure}[tb]
\includegraphics[scale=0.6]{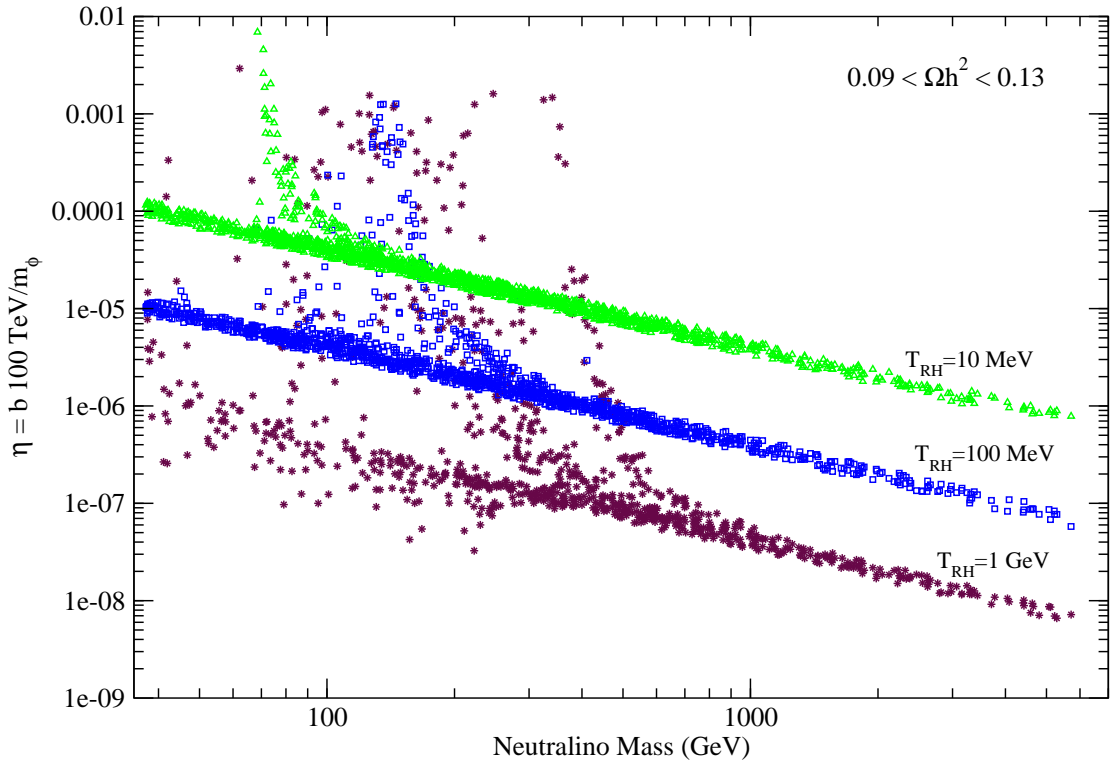}
\caption{$\eta$ as a function of the neutralino mass for models compatible with the WMAP range.}
\label{fig:wmapt}
\end{figure}

A question we have not directly addressed so far is  what  the parameter space compatible with the WMAP range is. Given the complexity of the parameter space involved ($10$ supersymmetric parameters plus $T_{RH}$ and $b$), it is not possible to answer this question in all its generality. What we can do is  fix some parameters and see if there are constraints on any others. In Fig.~\ref{fig:wmapt}, we show the value of $\eta$ required to obtain $0.09<\Omega_\chi h^2<0.13$  as a function of the neutralino mass, for three different values of the reheating temperature, $T_{RH}=1 ~\mathrm{GeV}, 100 ~\mathrm{MeV}, 10~\mathrm{MeV}$. 
Again here the straight bands proportional to $m_\chi^{-1}$ are due to non-thermal
production of neutralinos without chemical equilibrium (our case (3), eq.~\ref{nonthermal}).  For $T_{RH}=10 ~\mathrm{MeV}$, most neutralinos are produced non-thermally and the relic density is proportional to $\eta$ (Eq.~\ref{nonthermal}). Heavy neutralinos ($m_\chi>1 ~\mathrm{TeV}$) require $\eta<4\times 10^{-6}$  whereas the lightest neutralinos demand $\eta \simeq 10^{-4}$. From this band including most models, a small branch of wino-like neutralinos deviates (at $m_\chi\simeq 100~\mathrm{GeV}, \eta\simeq 4\times 10^{-5})$ and reaches values of $\eta$ as large as $10^{-2}$.  These models have a very small standard relic density and reach the required dark matter density range
through the solutions of our case (4), Eq.~\ref{eq:case4}, non-thermal production with chemical equilibrium, which require large values of $\eta$.
 For $T_{RH}=100 ~\mathrm{MeV}$, the figure is similar with the wino deviation occurring at larger masses ($m_\chi\simeq 250 ~\mathrm{GeV}$) and $\eta$ extending up to $10^{-3}$. For $T_{RH}=1 ~\mathrm{GeV}$, on the other hand, two different branches corresponding to winos and higgsinos can be distinguished. Moreover, some points lying below the non-thermal production band are also present.

\section{Conclusions}

In cosmologies with a decaying scalar field, the dark matter bound on supersymmetric models can almost always be satisfied. In fact, by adjusting $T_{RH}$ and $\eta$ the relic density can always be brought into the observed range except for
neutralinos with
$\Omega_{\rm std} \lesssim 10^{-5} (100 {\rm GeV}/m_\chi)$.

Models that in the standard cosmological scenario have a relic density above the WMAP range, as is generically the case for bino-like neutralinos, can be rescued by decreasing the reheating temperature of the Universe  (either just suppressing the thermal production, if $T_{RH}$ is below but close to the standard freeze-out temperature, or producing them non-thermally with a non-unique combination of $T_{RH}$ and $\eta$
 for lower $T_{RH}$ values). Models that in the standard scenario have a relic density below the WMAP range, as for higgsino- and wino-like neutralinos, can be saved with an appropriate and non-unique combination of $T_{RH}$ and $\eta$. 
 
 For dark matter detection purposes, the ideal neutralino -- i.e. one that
accounts for all of the dark matter content of the Universe,
has a relatively small mass, so that its number density is  large, and 
has large interaction and annihilation cross sections --
 is an oddity within the standard cosmological scenario. A light neutralino with a large interaction cross section typically yields a relic density smaller than the dark matter density. Thus, within  the standard cosmological  scenario, we are esentially led to two possibilities: a light neutralino with a small annihilation cross section, such as a bino-like neutralino; or a heavy neutralino ($m_{\chi}>1~\mathrm{TeV}$) with a large annihilation cross section, such as a higgsino or wino-like neutralino. If neutralinos are produced non-thermally, as is possible in low reheating temperature scenarios, the neutralino relic density is determined by the physics at the high scale, which
 determines  $T_{RH}$ and $\eta$, and not by the annihilation cross section. Therefore, by adjusting $T_{RH}$ and $\eta$, it becomes  possible for a light neutralino with a large annhihilation cross section to account entirely for the dark matter of the Universe. Ultra-heavy neutralinos, in the TeV mass range, become good dark matter candidates too.

\begin{figure}[tb]
\includegraphics[scale=0.45]{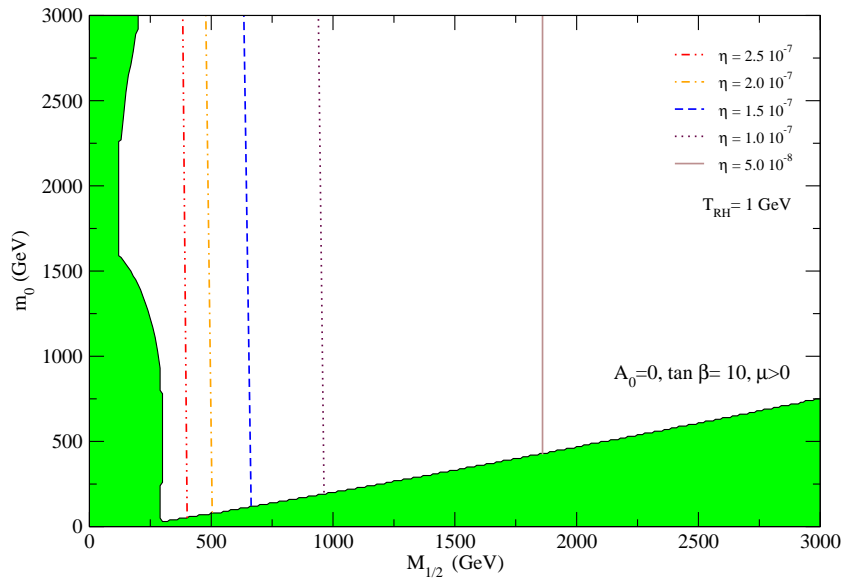}\\
\caption{The lines show the values of $m_0$ and $M_{1/2}$ for which the relic neutralino density
has the central value of the range imposed by WMAP bounds, namely $\Omega_{\rm cdm} h^2 =0.11$ for different values of $\eta$. Minimal SUGRA model with
$A_0=0$, $\tan\beta = 10$ and $\mu>0$.}
\label{fig:msugra}
\end{figure}

 For example, the usual narrow corridors in $m_0$, $M_{1/2}$ space of good dark matter neutralino candidates  in mSUGRA models are replaced by other narrow corridors, which, however, depend on the physics at the high scale that determines $T_{RH}$ and $\eta= b/m_\phi$. The $m_0$, $M_{1/2}$ plane for  a mSUGRA model with
$A_0=0$, $\tan\beta = 10$ and $\mu>0$ is shown in in Fig.~\ref{fig:msugra}. The gray (green) area is forbidden by experimental bound. The almost vertical lines in the figure show where neutralinos have $\Omega_{\rm cdm} h^2 =0.11$, the central value of the range imposed by WMAP bounds, for $T_{\rm RH}=$ 1GeV and different values of $\eta$. We see that in all points in the $m_0$, $M_{1/2}$ space in Fig.~\ref{fig:msugra} neutralinos can account for all of the cold dark matter, with appropriate
values of $T_{RH}$ and $\eta$.
In other words, in low reheating temperature scenarios with a decaying scalar field, cosmological data constrain models at the inflationary or Planck scale which determine $T_{RH}$ and $\eta$.

Future accelerator or dark matter detection experiments might find a neutralino in a region of the supersymmetric parameter space where its standard relic density is larger than the observed cold dark matter density.  This would tell us that the history
of the Universe before Big Bang Nucleosynthesis does not follow  the standard
cosmological assumptions, and would point towards non-standard scenarios, such as those with low reheating temperature studied here.

\vspace{0.3cm}
{\bf Acknowledgments}

This work was supported in part by the US Department of Energy Grant
DE-FG03-91ER40662, Task C and NASA grants NAG5-13399  and ATP03-0000-0057
at UCLA, and NFS
grant PHY-0456825 at the University of Utah. We thank A. Masiero and S. Nussinov
 for helpful discussions, O. Dor\'e for referring us to the website in Ref.~\cite{wmap2}, and the Aspen Center for Physics, where part of this work was done.

\end{document}